\documentclass[aps,prl,twocolumn,superscriptaddress,floatfix]{revtex4-2}
\usepackage{amssymb}
\usepackage{physics}
\usepackage{siunitx}
\usepackage{graphicx}
\usepackage{amsmath}
\usepackage{amsthm}
\usepackage{amsfonts}
\usepackage[T1]{fontenc}
\usepackage{array}
\usepackage{multirow}
\usepackage{color}
\usepackage{esint}
\usepackage{bm}
\usepackage{color}
\usepackage{bbm}
\usepackage{hyperref}
\usepackage{babel}

\newcommand{\todo}[1]{{\color{cyan} {#1}}}

\newcommand{\lettersection}[1]{\emph{\todo{#1}}.---}

\usepackage{hyperref}
\hypersetup
{	colorlinks,%
	citecolor=green,%
	linkcolor=red,%
	urlcolor=blue%
}
\allowdisplaybreaks[4]
\begin{document}

\global\long\def\id{\mathbbm{1}}
\global\long\def\ui{\mathbbm{i}}
\global\long\def\ud{\mathrm{d}}

\title{Thermal ionization of  impurity-bound quasiholes in the fractional quantum Hall effect}

\author{Ke Huang}
\affiliation{Department of Physics, City University of Hong Kong, Kowloon, Hong Kong SAR, China}

\author{Sankar Das Sarma}
\affiliation{Condensed Matter Theory Center and Joint Quantum Institute, University of Maryland, College Park, Maryland 20742, USA}

\author{Xiao Li}
\email{xiao.li@cityu.edu.hk}
\affiliation{Department of Physics, City University of Hong Kong, Kowloon, Hong Kong SAR, China}

\date{\today}

\begin{abstract}
We study the interplay between a Coulomb impurity and quasiholes in a fractional quantum Hall (FQH) state at finite temperatures. While a repulsive impurity can pin a quasihole and stabilize the FQH state, an attractive impurity cannot bind quasiholes. 
We demonstrate that at finite temperatures, a quasihole can be thermally ionized from a repulsive impurity, resulting in an ionization phase transition.  
We propose an experimental setup using exciton sensing to detect such a thermal ionization of quasiholes.
\end{abstract}

\maketitle

\lettersection{Introduction}
The fractional quantum Hall effect (FQHE) is a paradigm of strongly correlated electron systems~\cite{Tsui1982}, where the electron liquid forms a topologically ordered state~\cite{Wen1990,Wen1995}. 
In the FQH state, low-energy excitations known as quasiholes and quasiparticles are topological excitations that carry fractional charge and obey anyonic statistics~\cite{Laughlin1983,Arovas1984}. 
These excitations are central to understanding FQH physics and, due to their unique braiding statistics, in some situations provide a potential platform for topological quantum computation~\cite{Kitaev2003,DasSarma2005,Nayak2008}. 

Although FQHE is undoubtedly one of the most significant physics discoveries of the last 50 years, introducing fundamental concepts such as anyonic quasiparticles that obey fractional statistics and exhibit topological order, there are aspects of its physics that have not yet been thoroughly explored. 
One such aspect is the interplay between the FQHE and disorder. 
It is known that disorder suppresses FQHE as it is observed only in relatively clean samples with high mobility. 
On the other hand, however, the existence of the FQH quantized plateau itself depends on having some disorder so that the chemical potential can move smoothly through the gap, producing a finite plateau width. 
Thus, disorder plays a dual role, both allowing the quantization to manifest itself and also hindering the phenomenon, as the FQHE disappears in highly disordered samples. 
Even in very clean samples, the FQHE eventually disappears at some low fillings of the lowest Landau level, giving way to a highly resistive insulating phase, where disorder plays a key role. 
In the current work, we show that charged impurities (random quenched charged impurities are known to be the primary source for disorder in FQH samples) may lead to an unexpected and counterintuitive thermal ionization of quasiholes in FQHE, which should be experimentally observable. 
The binding of FQHE quasihole excitations to impurities plays a key role in the physics we predict. 

Although quasiholes are intrinsic excitations of the FQHE~\cite{Laughlin1983}, their observation often relies on disorder inherent to the sample. 
For example, Coulomb impurities in an FQHE provide local potentials that can give rise to many interesting properties~\cite{Zhang1985,Rezayi1985}. 
In particular, they can pin quasiholes in space, effectively trapping them for observation~\cite{Trugman1985}. 
This localization is critical for local probe techniques, including scanning tunneling microscopy~\cite{Feldman2009} and single-electron transistors~\cite{Martin2004}, which enable the direct detection and manipulation of fractional quasiparticles.
Quasihole localization by impurities is often a direct or indirect operational mechanism in FQHE experimental phenomena.

However, the binding of a quasihole to an impurity is not guaranteed at finite temperatures, where thermal fluctuations can overcome the binding energy and lead to the ionization of the quasihole. 
Therefore, braiding and manipulating quasiholes in the presence of impurities at finite temperatures becomes challenging. 
In fact, the stability of the quasihole trapped by the impurity is determined by its free energy $F=E-TS$ at finite temperatures. 
As a free quasihole has much greater entropy than a localized one, the quasihole may be released (i.e., thermally ionized) at $T_e>\delta E/\delta S$, where $\delta E$ and $\delta S$ are the energy penalty and the entropy gain of the excitation, respectively. 
Therefore, the ionization of the quasihole at finite temperatures could happen if $T_e$ is smaller than the FQH temperature scale $T_\text{FQH}$, beyond which the FQH state is destroyed by higher-energy excitations. 
Therefore, a careful analysis of the competition between energy and entropy is necessary to understand the behavior of quasiholes in the presence of impurities at finite temperatures.

In this work, we study this problem and characterize the possible thermal ionization process of quasiholes in a $\nu=1/3$ FQH state. 
We are interested in the interplay between quasiholes and impurities at zero and finite temperatures. 
At zero temperature, a repulsive impurity attracts and localizes the quasihole before additional quasiparticle-quasihole pairs are created. 
By contrast, an attractive impurity cannot trap quasiholes, and quasiholes remain itinerant. 
To begin with, we benchmark our method by characterizing the FQH state at exact $1/3$ filling in the presence of a Coulomb impurity using both the many-body energy spectrum and the particle entanglement spectrum (PES)~\cite{Sterdyniak2011,Regnault2011} at finite temperatures. 
We then consider a system with one quasihole and demonstrate that a repulsive impurity can pin the quasihole at zero temperature, while an attractive impurity cannot. 
We explicitly distinguish between the regimes of pinned versus free quasiholes using signatures in the PES and show that the corresponding energy and entanglement gaps have distinct dependences on impurity strength and temperature. 
Finally, we propose an experimental setup using interlayer excitons in transition metal dichalcogenide (TMD) heterobilayers as a quantum sensor to detect the thermal ionization of quasiholes in the FQH layer. 
In particular, we show that the exciton photoluminescence peak should blue-shift with increasing the temperature of the FQH layer only for a repulsive impurity configuration, while remaining nearly unchanged for an attractive impurity configuration, directly manifesting the thermal ionization process.

\begin{figure*}[t]
	\center
	\includegraphics[width=0.9\textwidth]{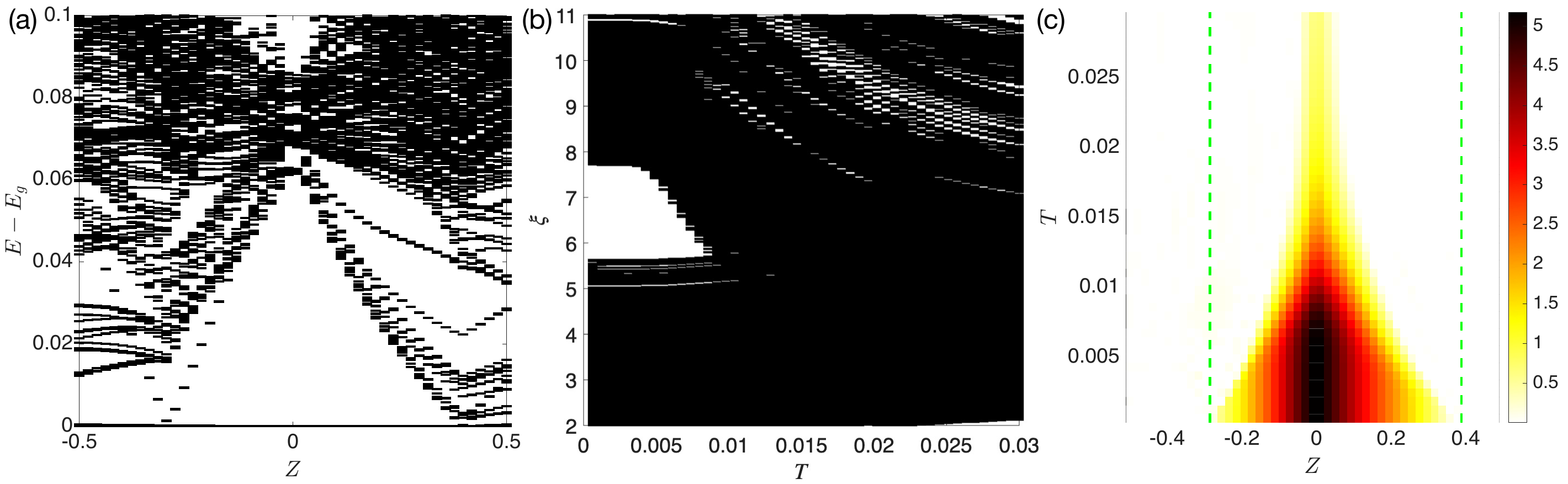}
	\caption{\label{Fig:no_qh}Calculation in the $N_\phi=21$ system with $N=7$ particles. (a) Energy spectrum as a function of the impurity charge $Z$. $E_g$ is the ground state energy. For $-0.3<Z<0.4$, the ground state is almost three-fold degenerate. (b) Particle entanglement spectrum by retaining $N_a=3$ particles at finite temperatures for $Z=0.2$. There are 637 states below the entanglement gap, consistent with the generalized Pauli principle for the system of 21 orbitals. 
	(c) Phase diagram of the FQH entanglement gap for a system of 21 orbitals. The two green vertical lines indicate where the energy gap vanishes at zero temperature. The color bar is normalized by the gap of the PES at zero temperature.
	}
\end{figure*}

\lettersection{The model}
We start by introducing the model Hamiltonian. 
Our model consists of a Coulomb impurity in the FQH state in a toroidal geometry. 
A torus is defined by imposing periodic boundary conditions on a parallelogram spanned by $\vb{L}_1$ and $\vb{L}_2$. 
Here, we take a square torus with $\abs{\vb{L}_1}=\abs{\vb{L}_2}$ and $\vb{L}_1\perp \vb{L}_2$. 
In the presence of a uniform perpendicular magnetic field, free electrons on a plane are discretized into Landau levels because of the non-commutativity of the canonical momenta, $[\pi_a,\pi_b]=i\epsilon_{ab}\hbar^2/l_B^2$, where $l_B=\sqrt{\hbar/(eB)}$ is the magnetic length.
The magnetic flux threaded through the torus is quantized and satisfies $B\abs{\vb{L}_1\cp \vb{L}_2}=N_\phi \phi_0$, where $N_\phi$ is an integer, and $\phi_0=h/e$ is the magnetic flux quantum. Consequently, each LL is $N_\phi$-fold degenerate.

If $\nu\leq1$ and Landau level mixing is negligible (as is appropriate in the high magnetic field FQHE limit), we can project the full Hamiltonian onto the lowest Landau level (LLL) and write the effective Hamiltonian as 
\begin{align}
	H = \mathcal{P}_{\text{LLL}}\bigl(V_{\text{int}} + Z V_{\text{imp}}\bigr) \mathcal{P}_{\text{LLL}},
\end{align}
where $\mathcal{P}_{\text{LLL}}$ is the LLL projector, $V_{\text{int}}$ describes the electron–electron interaction, and $Z V_{\text{imp}}$ is the impurity potential generated by a point charge $-Ze$. 
Therefore, the impurity is repulsive for $Z>0$ and attractive for $Z<0$. 
Throughout this work, we use the bare Coulomb interaction $e^2/(4\pi\epsilon r)$ and measure all energies and temperatures in units of $e^2/(4\pi\epsilon l_B)$. We elaborate on the detailed construction of the Hamiltonian in the supplemental material (SM)~\cite{supplement}.

\lettersection{The exact $1/3$ FQH state with a Coulomb impurity}
As a benchmark, we first consider the phase transition of the FQH state at exact $\nu=1/3$ filling. 
Specifically, we consider $N_\phi=21$ and $N=7$. 
The incompressible FQH state can sustain a finite impurity strength $Z$ before being destroyed by the quasiparticle-quasihole pairs created from the impurity site. 
Consequently, the FQHE is robust against a finite range of $Z$, and the FQH gap remains open for $-0.3<Z<0.4$, as shown in Fig.~\ref{Fig:no_qh}(a). 
For $Z>0.4$ and $Z<-0.3$, we demonstrate in the SM that the ground state is adiabatically connected to the magneto-roton mode~\cite{supplement}, implying that a strong enough Coulomb impurity creates a quasihole-quasiparticle pair, which forms the magneto-roton mode. 
The quasihole is pinned for the repulsive case, while the quasiparticle is pinned for the attractive case. 
Thus, the contrasting behaviors for attractive and repulsive impurities reflect the intrinsic asymmetry between quasiholes and quasiparticles in the FQHE.

Next, we study the finite-temperature properties and characterize the thermal state using the particle entanglement spectrum (PES)~\cite{Sterdyniak2011,Regnault2011}.
The PES is defined as the eigenvalue spectrum of
$\xi = -\ln \rho_A$, 
where the reduced density matrix $\rho_A=\Tr_B\rho$ is obtained from the full density matrix $\rho$ by partitioning the $N$ particles into two subsystems, $A$ and $B$, and tracing out subsystem $B$.
An FQH state is characterized by an entanglement gap in the PES, below which the number of states is given by the number of quasihole excitations. 
In Fig.~\ref{Fig:no_qh}(b), we show the PES as a function of temperature for a finite $Z = 0.2$. 
At low temperatures, there is an entanglement gap with a state counting consistent with the generalized Pauli principle~\cite{Bernevig2008}, which describes the quasihole excitations at $\nu=1/3$. 
The FQH gap protects this entanglement gap for a finite temperature range, but eventually vanishes at higher temperatures. 
We further show the phase diagram of the FQH entanglement gap in Fig.~\ref{Fig:no_qh}(c). 
For $-0.3<Z<0.4$, an entanglement gap is present at low temperatures, consistent with the energy spectrum in Fig.~\ref{Fig:no_qh}(a). 
Meanwhile, increasing the magnitude of $Z$ lowers the temperature at which the entanglement gap closes.
This is also consistent with the size of the energy gap in Fig.~\ref{Fig:no_qh}(a). 
For $Z>0.4$ and $Z<-0.3$, there is no entanglement gap even at zero temperature, indicating that the FQH ground state is overwhelmed by the magneto-roton mode. 
Finally, the FQH entanglement gap does not reemerge at finite temperatures, because the magneto-roton mode carries much larger entropy than the incompressible FQH state.

\begin{figure}[t]
	\center
	\includegraphics[width=0.47\textwidth]{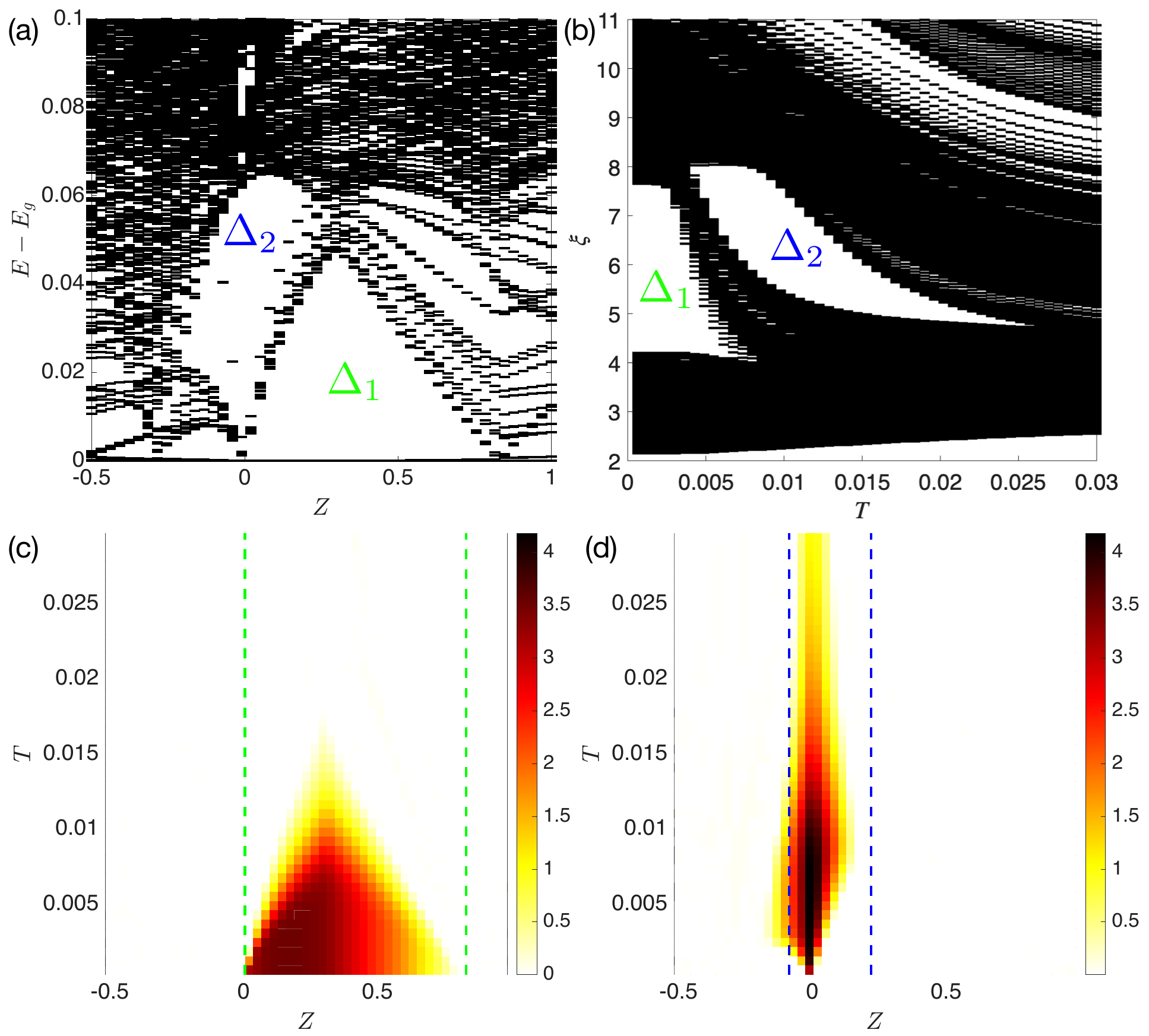}
	\caption{\label{Fig:qh}Calculation in the $N_\phi=22$ system with $N=7$ particles. 
	(a) Energy spectrum as a function of the impurity charge $Z$. $E_g$ is the ground state energy. 
	There are 22 states below the energy gap $\Delta_2$ (blue), corresponding to the number of free quasihole excitations, and three states below the energy gap $\Delta_1$ (green), corresponding to the FQH state with one pinned quasihole. 
	(b) Particle entanglement spectrum at finite temperatures for $Z=0.0816$. There are 637 states below the entanglement gap $\Delta_1$, consistent with the generalized Pauli principle for a system of 21 orbitals, and 770 states below the entanglement gap $\Delta_2$, consistent with the generalized Pauli principle for a system of 22 orbitals. 
	(c) Phase diagram of entanglement gap $\Delta_1$. The green vertical lines indicate where the energy gap $\Delta_1$ vanishes at zero temperature. 
	(d) Phase diagram of entanglement gap $\Delta_2$. 
	The blue vertical lines indicate where the energy gap $\Delta_2$ vanishes at zero temperature.
	Here, the particle entanglement spectrum is obtained by retaining $N_a=3$ particles. 
	}
\end{figure}

\lettersection{Thermal ionization of a quasihole}
Having analyzed the system at exact $1/3$ filling, we now turn to the case with one quasihole. 
We study a system with $N_\phi=22$ and $N=7$ to numerically verify the intuitive ionization process for a $\nu=1/3$ FQH state with one quasihole. 
We show the energy spectrum as a function of $Z$ in Fig.~\ref{Fig:qh}(a). 
Without the impurity, the ground state of the system is exactly $22$-fold degenerate, above which appears an energy gap [labeled as $\Delta_2$ in Fig.~\ref{Fig:qh}(a)] of a similar size as the FQH gap at the exact $\nu=1/3$. 
As $Z$ becomes nonzero, the energy gap $\Delta_2$ gradually closes; however, its behavior is asymmetric for attractive and repulsive impurities. 
On the attractive side, only the free–quasihole gap $\Delta_2$ is visible, and no additional sizable gap develops. 
In contrast, on the repulsive side, both the free–quasihole gap $\Delta_2$ and a second, equally pronounced gap $\Delta_1$ appear. 
Moreover, the ground state below the gap $\Delta_1$ is almost threefold degenerate. 
Such a degeneracy is consistent with the case where a pinned quasihole removes one available orbital, thereby reducing the effective system size by one. 
Therefore, there are seven particles in an effective system of 21 orbitals, resulting in an effective 1/3 filling and a $\nu = 1/3$ FQH state. 
These considerations show that the quasihole is now pinned by the impurity.

The PES result further corroborates this conclusion. 
The PES can distinguish between pinned quasiholes and free quasiholes because the entanglement gap counts quasihole excitations, which are sensitive to the system size. 
The resulting state counting, therefore, reflects both the reduced system size and the number of pinned quasiholes.
This expectation is indeed verified in Fig.~\ref{Fig:qh}(b), where we retain $N_a=3$ particles and calculate the PES for $Z=0.0816$ as a function of temperature. 
At low temperatures, the entanglement gap $\Delta_1$ [see Fig.~\ref{Fig:qh}(b)] has 637 states below it and follows the state counting of a system of 21 orbitals, rather than 22 orbitals. 
Nevertheless, the entanglement gap $\Delta_1$ closes as the temperature increases. 
Meanwhile, the entanglement gap $\Delta_2$ emerges in the PES, with a state counting of 770, corresponding to a system of 22 orbitals. 
The crossover to the entanglement gap $\Delta_2$ implies a free quasihole at intermediate temperatures and confirms the expected thermal ionization. 
To summarize, both energy and entanglement gaps $\Delta_1$ correspond to a pinned quasihole, while energy and entanglement gaps $\Delta_2$ correspond to a free quasihole.

We also compute the phase diagram using the two entanglement gaps. 
The phase diagram of the entanglement gap $\Delta_1$ in Fig.~\ref{Fig:qh}(c) only opens for a repulsive impurity, as expected.
Except that, the entanglement gap $\Delta_1$ resembles the phase diagram at the exact $\nu=1/3$ in Fig.~\ref{Fig:no_qh}(c), because they both originate from the incompressible FQH states. Notably, as the entropy of the ground state is too low to play a role, the entanglement gap decreases monotonically as the temperature decreases. 
On the contrary, a free quasihole has a much higher entropy and leads to a distinct phase diagram for entanglement gap $\Delta_2$, as shown in Fig.~\ref{Fig:qh}(d). 
Though the entanglement gap vanishes at low temperatures, it can appear at finite temperatures, where the entropy of the free quasihole dominates. This competition between energy and entropy enables the thermal ionization of quasiholes.

\begin{figure}[t]
	\center
	\includegraphics[width=0.7\columnwidth]{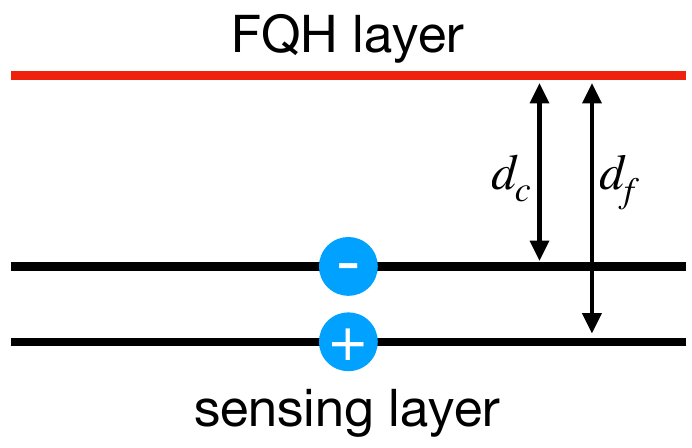}
	\caption{\label{Fig:exciton}Schematics of the experimental proposal using an interlayer exciton.	}
\end{figure}

\lettersection{Experimental proposal}
The above analysis employed Coulomb disorder; however, the exact thermal ionization mechanism applies to any localized impurity, whether attractive or repulsive. 
A direct method for detecting ionization is interlayer-exciton sensing, which has recently demonstrated high sensitivity to correlated electronic states in two-dimensional materials~\cite{Miao2021,Cai2023,Wang2023,Xia2024}. 
As shown in Fig.~\ref{Fig:exciton}, we place a type-II TMD heterobilayer adjacent to the FQH layer; the electron and hole reside in different monolayers and form an interlayer exciton~\cite{Rivera2018,Jiang2021}. When pinned by disorder, a single exciton acts as a static, charge-neutral dipolar impurity with potential
\begin{align}
	V(r)=\pm \dfrac{e^2}{4\pi \epsilon}\qty[(r^2+d_c^2)^{-1/2}-(r^2+d_f^2)^{-1/2}],
\end{align}
where $d_c$ ($d_f$) is the distance from the FQH layer to the nearer (farther) TMD layer. The $+$ ($-$) sign corresponds to the exciton’s negative charge residing in the nearer (farther) layer, yielding an overall repulsive (attractive) impurity; flipping the heterobilayer swaps the sign, while varying $d_c$ and $d_f$ tunes the coupling strength without changing the charge. In the repulsive configuration, an exciton can lower its energy by binding a quasihole at low temperature~\cite{Mostaan2025,Wagner2025}; upon heating, the quasihole ionizes, and the exciton energy increases. In contrast, quasihole binding is absent in the attractive configuration. 
This leads to a clear optical signature: the exciton photoluminescence peak should blue-shift with increasing the temperature of the FQH layer only for the repulsive orientation, while remaining nearly unchanged for the attractive orientation.
The observation of this predicted blueshift would directly confirm the reentrance behavior shown in Fig.~\ref{Fig:qh}, arising from the competition between energy and entropy.

\lettersection{Conclusion}
In this work, we have demonstrated that a single Coulomb impurity in a $\nu=1/3$ FQH fluid can drive a genuine thermal ionization of a quasihole. 
At zero temperature, a repulsive impurity binds and pins a quasihole, generating an incompressible state that is adiabatically connected to the Laughlin phase on an effectively reduced number of orbitals. 
By using both the many-body energy spectrum and the PES on the torus, we explicitly distinguished between the regimes of pinned versus free quasiholes. We demonstrated that the corresponding energy and entanglement gaps, $\Delta_1$ and $\Delta_2$, exhibit distinct dependencies on the impurity strength and temperature.

The essential ingredient underlying the observed ionization process is the entropy difference between a localized and a delocalized quasihole. 
While the FQH state with a pinned quasihole is energetically favorable at low temperatures, the free quasihole carries a much larger configurational entropy, which can compensate for the energy cost at intermediate temperatures. 
This competition leads to a finite-temperature regime where the entanglement structure switches from that of a pinned quasihole (effective $N_\phi=21$) to that of a free quasihole (true $N_\phi=22$).  
Our phase diagrams of the entanglement gaps quantitatively capture this entropy-driven crossover, emphasizing the intrinsic asymmetry between repulsive and attractive impurities and the explicit competition between energy and entropy.  

Beyond Coulomb impurities, the exact thermally induced ionization mechanism should apply to a broad class of localized impurity potentials that couple to the charge sector of the FQHE. 
We have proposed a realistic experimental platform based on interlayer excitons in TMD heterobilayers capacitively coupled to an FQH layer. 
In this setup, a localized exciton acts as a tunable attractive or repulsive impurity, whose effective strength is controlled geometrically. The binding or unbinding of a quasihole manifests as a characteristic temperature-dependent shift in the photoluminescence peak. 
Our results thus provide a concrete route to directly probe fractionalized quasiholes, their impurity binding, and their entropy-driven ionization using exciton spectroscopy in moiré and semiconductor quantum Hall heterostructures. 
Such direct observations of fractionalized quasiholes would be a significant and exciting new development in the field.

\lettersection{Acknowledgement}
K.H. and X.L. are supported by the Research Grants Council of Hong Kong (Grants No. CityU 11300421, CityU 11304823, CityU 11312825, and C7012-21G) and City University of Hong Kong (Project No.~9610428). K.H. is also supported by the Hong Kong PhD Fellowship Scheme. 
S.D.S. is supported by the Laboratory for Physical Sciences through the Condensed Matter Theory Center (CMTC) at the University of Maryland.

\bibliography{FQH_v3.bib}

\end{document}


\global\long\def\id{\mathbbm{1}}
\global\long\def\ui{\mathbbm{i}}
\global\long\def\ud{\mathrm{d}}

\title{Supplemental Material for ``Thermal ionization of  impurity-bound quasiholes in the fractional quantum Hall effect''}

\author{Ke Huang}
\affiliation{Department of Physics, City University of Hong Kong, Kowloon, Hong Kong SAR, China}

\author{Sankar Das Sarma}
\affiliation{Condensed Matter Theory Center and Joint Quantum Institute, University of Maryland, College Park, Maryland 20742, USA}
\affiliation{Kavli Institute for Theoretical Physics, University of California, Santa Barbara, California 93106, USA}

\author{Xiao Li}
\email{xiao.li@cityu.edu.hk}
\affiliation{Department of Physics, City University of Hong Kong, Kowloon, Hong Kong SAR, China}

\date{\today}

\maketitle

\setcounter{equation}{0} 
\setcounter{figure}{0}
\setcounter{table}{0} 
\renewcommand{\theparagraph}{\bf}
\renewcommand{\thefigure}{S\arabic{figure}}
\renewcommand{\theequation}{S\arabic{equation}}
\renewcommand\labelenumi{(\theenumi)}

\maketitle

\section{Lowest Landau level on a torus}

In this section, we adopt an abegrabic approach to construct the fractional quantum Hall system using toroidal geometry.
The Hilbert space of the lowest Landau level on a torus can be constructed by imposing periodic conditions on the Hilbert space on a plane. In the presence of a uniform perpendicular magnetic field, free electrons on a plane are tdiscretized into LLs because of the non-commutativity of the canonical momenta, $[\pi_a,\pi_b]=i\epsilon_{ab}\hbar^2/l_B^2$, where $l_B=\sqrt{\hbar/(eB)}$ is the magnetic length. We set $\hbar=1$ and $l_B=1$ as the unit of length. The LLs can be constructed by a pair of bosonic creation and annihilation operators defined by $a=(\pi_x+i\pi_y)/\sqrt{2}$. Nonetheless, the creation operator does not resolve the degeneracy of the LLs. Hence, we consider the guiding center momenta $\vb*Q=\vb*\pi-\bar{\vb*r}$, where we define the Hodge dual $\bar{x}_a=\epsilon_{ab} x_b$. The guiding center momenta satisfy
\begin{align}
	[r_a,Q_b]=i\delta_{ab},\ \  [ \pi_a,Q_b]=0,\ \ [Q_a,Q_b]&=-i\epsilon_{ab}.
\end{align}
The first commutation indicates that the magnetic translation $T(\vb* x)=e^{i\vb* x\vdot \vb* Q}$ translates the position operator as the ordinary translation operator. The second commutation implies that $T(\vb* x)$ commutes with the creation and annihilation operators, resolving the degeneracy of the LLs. However, it is not always possible to construct the common eigenstates of two magnetic translations in the LLL because of the third commutation. Particularly, we have
\begin{align}
	T(\vb* R_1)T(\vb* R_2)=T(\vb*R_2)T(\vb* R_1)e^{i\vb* R_1 \cross \vb* R_2},
\end{align}
where we define $\vb* x \cross \vb* y=\epsilon_{ab} x_ay_b$. There exists a set of common eigenstates in the LLL if $\vb* R_1 \cross \vb* R_2=2\pi$, and the momentum of a common eigenstate is defined as
\begin{align}
	T(\vb* R_1)\ket{\psi}=e^{i\vb*k\vdot\vb*R_1}\ket{\psi},\quad T(\vb* R_2)\ket{\psi}=e^{i\vb*k\vdot\vb*R_2}\ket{\psi}.
\end{align}
Moreover, $T(\vb* x)$ also changes the momentum of a state by
\begin{align}
	T(\vb* R_\alpha)T(\vb* x)\ket{\psi}
	=e^{i\vb* R_\alpha \cross \vb* x}T(\vb* x)T(\vb* R_\alpha)\ket{\psi}
	=e^{i(\vb*k+\bar{\vb*x})\vdot\vb*R_\alpha}T(\vb* x)\ket{\psi}.
\end{align}
Consequently, all the states in the LLL can be generated by the states with zero momentum through
\begin{align}\label{Eq:Basis}
	\ket{\vb*k,\lambda}=T(-\bar{\vb*k})\ket{\vb*0,\lambda},
\end{align}
where $\lambda$ labels the basis states of an irreducible representation of the little group that does not change the momentum of a state. $T(-\bar{\vb*q})$ does not change the momentum if and only if $\vb*q=m\vb*G_1+n \vb*G_2$, where $m,n$ are two integers, and $\vb G_1=\bar{\vb*R}_2$ and $\vb G_2=-\bar{\vb*R}_1$ are the two reciprocal vectors of $\vb*R_1$ and $\vb*R_2$. Thus, the little group is an Abelian group, given by $\{T(m\vb*R_1+n\vb*R_2):\ m,n\in\mathbb{Z}\}$, and the irreducible representation is one dimensional. Henceforth, we omit the $\lambda$ index in Eq.~\eqref{Eq:Basis}.

The Hilbert space of the LLL on a torus spanned by $\vb* L_1$ and $\vb* L_2$ is the subspace that satisfies $T(\vb*L_1)\ket{\psi}=T(\vb*L_2)\ket{\psi}=\ket{\psi}$. Because of the non-commutativity $T(\vb*L_1)$ and $T(\vb*L_2)$, such a subspace is nonempty if and only if $\vb* L_1 \cross \vb* L_2=2\pi N_\phi$ for an integer $N_\phi$, meaning that the magnetic flux threaded through the torus must be an integer multiple of the magnetic flux quantum $\phi_0=h/e$. We have the freedom to choose $\vb*R_1$ and $\vb*R_2$, and a convenient choice is $\vb*R_1=\vb*L_1$ and $\vb*R_2=\vb*L_2/N_\phi$. The periodic condition translates to the constraint that $\vb*k$ must be integer multiples of $\vb*G_2/N_\phi$, and therefore, the basis states can be denoted by $\ket{n}:=\ket{n \vb*G_2/N_\phi}$ for all $n\in\mathbb{Z}$ with periodicity $\ket{n+N_\phi}=\ket{n}$.

\section{Coulomb interaction and impurity}

The Hamiltonian considered here is composed of two parts.
A translationally invariant interaction in the LLL can be expressed in the following form,
\begin{align}
	H_{\text{int}}&=\frac{1}{2A}\sum_{\vb*q,\vb*k,\vb*k'}V(q)
	\mel{\vb*k+\vb*q}{e^{i\vb* q\vdot \vb* r}}{\vb*k}
	\mel{\vb*k'-\vb*q}{e^{i\vb* q\vdot \vb* r}}{\vb*k'} c_{\vb*k+\vb*q}^\dag c_{\vb*k'-\vb*q}^\dag c_{\vb*k'}c_{\vb*k},
\end{align}
where $A$ is the area of the system, $c_{\vb*k}$ annihilates a state in the LLL with momentum $\vb*k$, and $V(q)=e^2/(4\pi\epsilon q)$ is the Coulomb potential. Limiting the interaction to the torus, we obtain
\begin{align}
	H_{\text{int}}&=\frac{1}{2N_\phi}\sum_{n,n'=0}^{N_\phi-1}\sum_{m=-\infty}^{+\infty}V_{mnn'}^{\text{int}} c_{n+m}^\dag c_{n'-m}^\dag c_{n'}c_{n},
\end{align}
where we define 
\begin{align}
	V_{mnn'}^{\text{int}}&=\sum_{m'=-\infty}^{+\infty}V(q_{mm'})\mel{n+m}{e^{i\vb* q_{mm'}\vdot \vb* r}}{n}
	\mel{n'-m}{e^{-i\vb* q_{mm'}\vdot \vb* r}}{n'}
\end{align}
with $\vb*q_{mm'}=m\vb*G_2/N_\phi+m'\vb*G_1$. A Coulomb impurity of charge $-Ze$ located at $\vb*r=\vb*0$ is given by
\begin{align}
	H_{\text{Coulomb}}&= Z\sum_{\vb*k,\vb*q}V(q)\mel{\vb*k+\vb*q}{e^{i\vb*q\vdot \vb*r}}{\vb*k}c^\dag_{\vb*k+\vb*q}c_{\vb*k}
\end{align}
which, on the torus, becomes
\begin{align}
	H_{\text{imp}}&=Z\sum_{n=0}^{N_\phi-1}\sum_{m=-\infty}^{+\infty}V_{nm}^{\text{imp}}c^\dag_{n+m}c_{n}
\end{align}
with
\begin{align}
	V_{nm}^{\text{imp}}=\sum_{m'=-\infty}^{+\infty}V(q_{mm'})\mel{n+m}{e^{i\vb*q_{mm'}\vdot\vb*r}}{n}.
\end{align}
All left to calculate is $\mel{n+m}{e^{i\vb*q_{mm'}\vdot\vb*r}}{n}$. Replacing $\vb*r$ with $\vb*\pi$ and $\vb*Q$, we obtain
\begin{align}
	\mel{n+m}{e^{i\vb*q_{mm'}\vdot\vb*r}}{n}
	=\mel{n+m}{e^{i\vb*q_{mm'}\vdot(\bar{\vb*Q}-\bar{\vb*\pi} ) }}{n}
	=e^{i\pi m'(2n+m)/N_\phi}e^{-q_{mm'}^2/4}.
\end{align}

\section{Spherical geometry and magneto-roton mode}

In this section, we briefly review the construct of the LLL for spherical geometry~\cite{Haldane1983}.
The LLs on a sphere of radius $R$ are generated by placing a magnetic monopole at the center of the sphere. Because of the rotational symmetry, each LL is a projective representation of SO(3). The magnetic flux $\Phi=4\pi R^2 B$ is quantized to an integer $2S$ of the flux quantum, and the total angular momentum of the LLL is $S$. The eigenstate with quantum number $L_z=m$ in the LLL is given by
\begin{align}
	\psi_m\propto u^{S+m}v^{S-m},
\end{align}
where $u=\cos(\theta/2)e^{i\phi/2}$,  $v=\sin(\theta/2)e^{-i\phi/2}$, and $\theta,\phi$ are the polar and azimuthal angle, respectively. The Coulomb interaction is $V(r)=1/(4\pi\epsilon r)$, where $r$ is the chord distance between two points on the sphere. By projecting it to the LLL, we obtain
\begin{align}
	H_{\text{int}}= \sum_{j}V_{Sj} \sum_{m} P_{jm}^\dag P_{jm},
\end{align}
where $V_j$ denotes pseudopotential for the Coulomb interaction, $P_{jm}=\sum_m C^{jm}_{Sm_1,Sm_2}c_{m_1} c_{m_2}$, $C^{jm}_{Sm_1,Sm_2}$ is the Clebsch-Gordan coeffficient, and $c_m$ is the annihilation operator for $\psi_m$. The pseudopotential can be calculated by
\begin{align}
	V_{Sj}&=\frac{\int_{0}^{\pi}\dd{\theta_1}\int_{0}^{\pi}\dd{\theta_2}\int_{0}^{2\pi}\dd{\phi_1}\int_{0}^{2\pi}\dd{\phi_2} \abs{r_1-r_2}^{-1}\abs{u_1v_2-u_2v_1}^{4S-2j}\abs{u_1}^{2j}\abs{u_2}^{2j}}{\int_{0}^{\pi}\dd{\theta_1}\int_{0}^{\pi}\dd{\theta_2}\int_{0}^{2\pi}\dd{\phi_1}\int_{0}^{2\pi}\dd{\phi_2} \abs{u_1v_2-u_2v_1}^{4S-2j}\abs{u_1}^{2j}\abs{u_2}^{2j}}\nonumber\\
	&=e^2(2\pi\epsilon \sqrt{S})^{-1}\binom{4S-2j}{2S-j}\binom{4S+2j+1}{2S+j+1}/\binom{4S+2}{2S+1}^2.
\end{align}
Due to the rotational symmetry, we place the Coulomb impurity at the North pole without loss of generality. The Coulomb impurity is thus given by
\begin{align}
	H_{\text{imp}}=Z\sum_m v_{Sm} c^\dag_m c_m,
\end{align}
where $v_{Sm}=\frac{(2S+1)e^2}{4\pi\epsilon\sqrt S}\binom{2S}{S+m}\int_0^1 2x^{2(S-m)}(1-x^2)^{2(S+m)}\dd{x}$. Without impurity, the total angular momentum is conserved, and each eigenvalue has an $L$-fold degeneracy. In the presence of a Coulomb impurity, $L$ is no longer a good quantum number, but $L_z$ remains conserved. Hence, compared with toroidal geometry, spherical geometry has the advantage of possessing a quantum number, which enables us to track eigenstates in their respective symmetry sector. However, the disadvantage of spherical geometry is that its particle entanglement spectrum lacks a global gap~\cite{Sterdyniak2011}, complicating the analysis at finite temperatures.

\begin{figure*}[t]
	\center
	\includegraphics[width=0.8\textwidth]{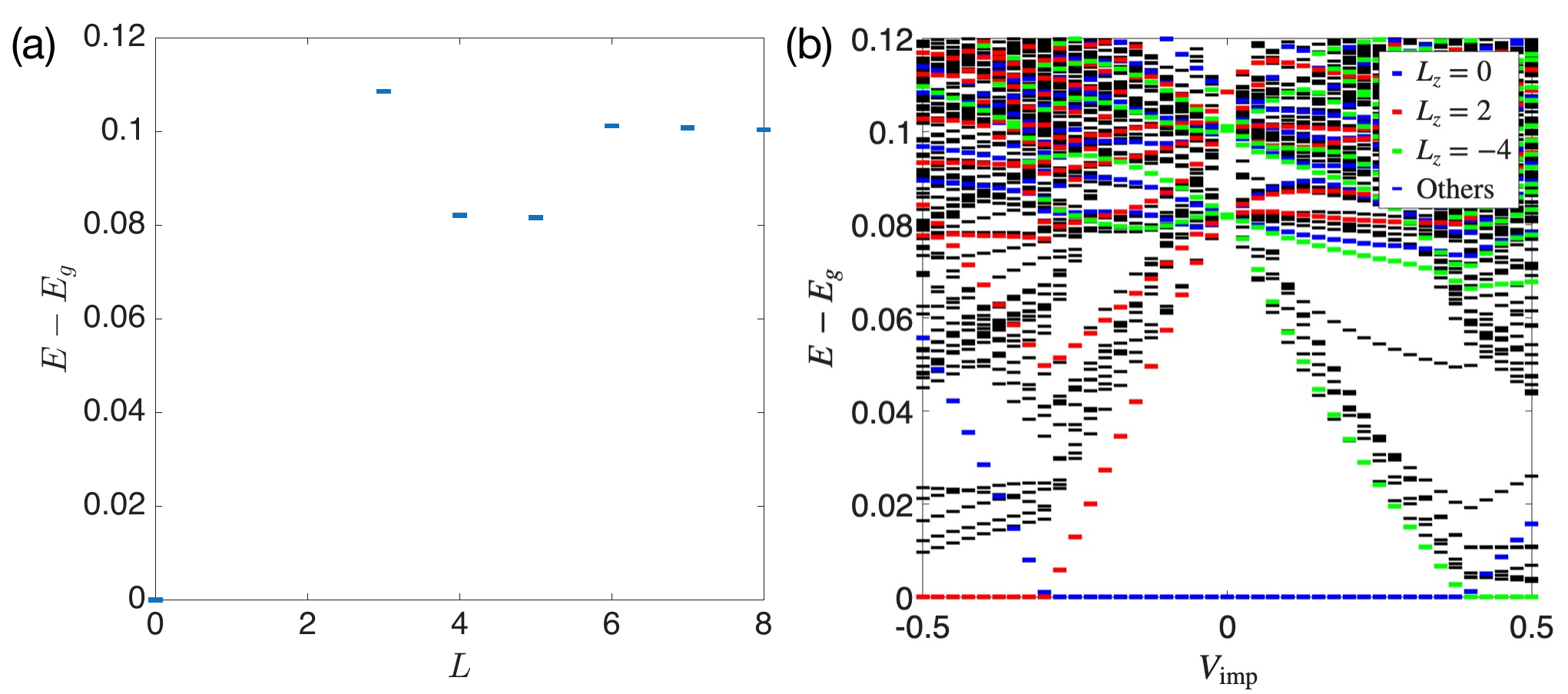}
	\caption{\label{Fig:sphere}(a) Angular momentum-resolved energy spectrum for the FQH state on a sphere. (b) Energy spectrum as a function of impurity strength $Z$. The colors label the quantum number $L_z$ of the eigenstate.  Here, the system size is $S=21/2$ with $N=8$ particles.
	}
\end{figure*}

In Fig.~\ref{Fig:sphere}(a), we calculate a system with $S=21/2$ and $N=8$ particles, corresponding to the $\nu=1/3$ fractional quantum Hall (FQH) state with no quasiholes. The two lowest energy states with $L=4$ and $L=5$ are identified as magneto-roton mode~\cite{Haldane1985,Girvin1986}. In Fig.~\ref{Fig:sphere}(b), we calculate the spectrum as a function of $Z$. The spectrum strikingly resembles that for a torus, because the magnetic length $l_B$ is much smaller than the system size. Although the geometry has a minimal effect on the shape of the spectrum, the degeneracy is sensitive to the geometry. Particularly, within $-0.3<Z<0.4$, the ground state is onefold for spherical geometry but threefold for toroidal geometry. Moreover, the spectrum also definitively shows that the magneto-roton mode with $L_z=2$ and $L_z=-4$ becomes the ground state for the strongly attractive and repulsive cases, respectively. 

\bibliographystyle{apsrev4-2}
\bibliography{FQH.bib}